\documentclass[11pt]{article}
\usepackage{amsmath,amssymb,epsf}

\def\ben{\begin{equation}}
\def\een{\end{equation}}

  \let\n=\nu

\let\C=\Chi

\def\nn{\nonumber} \def\bd{\begin{document}} \def\ed{\end{document}}
\def\ds{\documentstyle} \let\fr=\frac \let\bl=\bigl \let\br=\bigr
\let\Br=\Bigr \let\Bl=\Bigl
\let\bm=\bibitem
\let\na=\nabla
\let\pa=\partial \let\ov=\overline
\newcommand{\be}{\begin{equation}}
\newcommand{\ee}{\end{equation}}
\def\ba{\begin{array}}
\def\ea{\end{array}}
\def\ft#1#2{{\textstyle{\frac{\scriptstyle #1}{\scriptstyle #2}}}}
\def\fft#1#2{\frac{#1}{#2}}
\def\del{\partial}
\def\vp{\varphi}
\def\sst#1{{\scriptscriptstyle #1}}
\def\oneone{\rlap 1\mkern4mu{\rm l}}
\def\td{\tilde}
\def\wtd{\widetilde}
\def\ie{\rm i.e.\ }
\def\dalemb#1#2{{\vbox{\hrule height .#2pt
        \hbox{\vrule width.#2pt height#1pt \kern#1pt
                \vrule width.#2pt}
        \hrule height.#2pt}}}
\def\square{\mathord{\dalemb{6.8}{7}\hbox{\hskip1pt}}}
\newcommand{\ho}[1]{$\, ^{#1}$}
\newcommand{\hoch}[1]{$\, ^{#1}$}
\newcommand{\bea}{\begin{eqnarray}}
\newcommand{\eea}{\end{eqnarray}}
\newcommand{\ra}{\rightarrow}
\newcommand{\lra}{\longrightarrow}
\newcommand{\Lra}{\Leftrightarrow}
\newcommand{\ap}{\alpha^\prime}
\newcommand{\bp}{\tilde \beta^\prime}
\newcommand{\tr}{{\rm tr} }
\newcommand{\Tr}{{\rm Tr} }
\def\0{{\sst{(0)}}}
\def\1{{\sst{(1)}}}
\def\2{{\sst{(2)}}}
\def\3{{\sst{(3)}}}
\def\4{{\sst{(4)}}}
\def\5{{\sst{(5)}}}
\def\6{{\sst{(6)}}}
\def\7{{\sst{(7)}}}
\def\8{{\sst{(8)}}}
\def\n{{\sst{(n)}}}
\def\cA{{{\cal A}}}
\def\cB{{{\cal B}}}
\def\cF{{{\cal F}}}
\def\tV{\widetilde V}
\def\tW{\widetilde W}
\def\tH{\widetilde H}
\def\tE{\widetilde E}
\def\tF{\widetilde F}
\def\tA{\widetilde A}
\def\im{{{\rm i}}}
\def\tY{{{\wtd Y}}}
\def\ep{{\epsilon}}
\def\vep{{\varepsilon}}
\def\R{\rlap{\rm I}\mkern3mu{\rm R}}
\def\bD{{{\bar D}}}

\def\R{\rlap{\rm I}\mkern3mu{\rm R}}
\def\bD{{{\bar D}}}
\def\R{{{\mathbb R}}}
\def\C{{{\mathbb C}}}
\def\H{{{\mathbb H}}}
\def\CP{{{\mathbb C}{\mathbb P}}}
\def\RP{{{\mathbb R}{\mathbb P}}}
\def\Z{{{\mathbb Z}}}
\def\bA{{{\mathbb A}}}
\def\bB{{{\mathbb B}}}
\def\bC{{{\mathbb C}}}
\def\bD{{{\mathbb D}}}
\def\bE{{{\mathbb E}}}
\def\bZ{{{\mathbb Z}}}
\def\Re{{{\frak{Re}}}}
\def\Im{{{\frak{Im}}}}
\def\cosec{{\,\hbox{cosec}\,}}
\def\Gm{{\Gamma_{\!\! -}}}
\def\Gp{{\Gamma_{\!\! +}}}
\def\stan{{standard }}
\def\nonstan{{supernumerary }}

\begin{document}
\begin{flushright}
MIFP-05-03 \\
{\bf hep-th/0501213}\\
January\  2005
\end{flushright}

\begin{center}
 
{\large {\bf AdS pp-waves}}
 
\vspace{20pt}

Johannes Kerimo\hoch{1}

\vspace{20pt}

{\it George P. \&  Cynthia W. Mitchell Institute for Fundamental
Physics,\\
Texas A\&M University, College Station, TX 77843-4242, USA}

\vspace{40pt}
 
\underline{ABSTRACT}
\end{center}

		We obtain the pp-waves of $D=5$ and $D=4$ gauged supergravities
supported by $U(1)^3$ and $U(1)^4$ gauge field strengths respectively.
We show that generically these solutions preserve 1/4 of the
supersymmetry, but supernumerary supersymmetry can arise for
appropriately constrained harmonic functions associated with 
the pp-waves. In particular it implies that the solutions are
independent of the light-cone coordinate $x^+$.
We also obtain the pp-waves in the Freedman-Schwarz model.

{\vfill\leftline{}\vfill \vskip 10pt \footnoterule {\footnotesize
\hoch{1} Research supported in part by DOE grant
DE-FG03-95ER40917.
}


\newpage

\section{Introduction}

		The subject of pp-waves and their applications have been studied 
extensively. In particular, superstring theory is exactly solvable 
\cite{metsaev} on the backgrounds of the maximal supersymmetric pp-waves 
of type IIB \cite{bfhp} and M-theory \cite{kg}. This provides a 
rare example where the AdS/CFT duality \cite{malda,gkp,wit} can be 
tested \cite{bmn} beyond the supergravity approximation.

		When the integration constants associated with these solutions 
are left arbitrary the pp-waves generically preserve half of the 
supersymmetry. Additional supersymmetry (supernumerary supersymmetry) 
beyond the $\ft12$ can arise when the harmonic function is
constrained appropriately. 
For a discussion of pp-waves and their supersymmetry in M-theory and 
type IIB see \cite{clp1,clp2,gh,bena,mich1,mich2}.

		PP-waves on AdS background have also been studied.
The purely gravitational AdS pp-wave is given by
\be
ds^2=e^{2g\rho}(-4dx^+dx^- + H(dx^+)^2+dz_a^2)+d\rho^2\,,
\ee
where the cosmological constant is related to the gauge coupling constant $g$
as $\Lambda=-g^2$, and $H=H(x^+,\rho,z_a)$ is a harmonic function on the 
space of $z_a$ and $\rho$\,.
The integration constants of $H$ are therefore allowed to have an arbitrary 
dependence on $x^+$. If $H=0$ the metric describes pure AdS spacetime.
The pp-wave with the dependence $H(\rho)$ was constructed in four dimensions 
by Kaigorodov \cite{kaig} and its higher dimensional counterparts were 
obtained in \cite{clpboost}. Generalisations of the Kaigorodov metric 
to inhomogeneous solutions were obtained in \cite{siklos,ozsvath,gr}.

		The superimposing of a pp-wave on AdS spacetime can be viewed 
as performing an infinite boost on the boundary conformal field theory 
\cite{clpboost,bcr}. These solutions generically preserve $\ft14$ of
the maximum supersymmetry allowed by the AdS spacetime
\cite{clpboost,bcr}. The purely gravitational AdS pp-waves have 
in fact been shown to admit supernumerary supersymmetries \cite{kerimo}
for appropriately constrained $H$. 

		AdS pp-waves can also be supported by a field strength.
Their supersymmetry has been studied in \cite{gg,klemm1,klemm2,kerimo}. 
See also \cite{cai}. In the case of charged pp-waves of minimum gauged 
supergravities in $D=4$ and $D=5$, it was shown \cite{klemm2,kerimo}
that supernumerary supersymmetry can arise again for appropriately 
constrained harmonic functions. For pp-waves with $\ft12$ supersymmetry
in $D=3$ see \cite{deger}.
 
		In this paper we shall investigate pp-waves on AdS background 
further by studying the pp-waves of $D=5$ and $D=4$ gauged 
supergravities supported respectively by $U(1)^3$ and $U(1)^4$ 
gauge fields. We present a detailed analysis of the supersymmetry 
of these solutions. In particular, we show that supernumerary 
supersymmetry can arise beyond the usual $\ft14$.
We also study the pp-waves of the Freedman-Schwarz model.  
The supersymmetry enhancement obtained in this paper forces 
the solutions to be independent of the light-cone coordinate $x^+$.

	The paper is organised as follows. The pp-waves of gauged supergravities 
in five and four dimensions are studied in sections $2$ and $3$
respectively. We investigate the pp-waves of the Freedman-Schwarz model
in section $4$. In sections $5$ and $6$ we study the supersymmetry of 
the solutions in six and seven dimensions, respectively.

\section{PP-waves in five dimensions}
Our first example treats $D=5$ gauged supergravity truncated to 
the $U(1)^3$ subgroup of $SO(6)$. The bosonic sector of this truncated theory 
is described by the Lagrangian \cite{clp}
\be
e^{-1}{\cal L}_5=R-\ft12(\del\vec{\varphi})^2
+ 4g^2\sum_i X_i^{-1} - \ft14\sum_iX_i^{-2}(F^{\,i}_\2)^2
+ \ft14\epsilon^{\sst{MNPQR}}F_{\sst{MN}}^1
F_{\sst{PQ}}^2A_{\sst R}^3\,,
\ee
where $\vec{\varphi}=(\varphi_1,\varphi_2)$, and we write
\bea
&&X_i=e^{-\ft12\vec{a}_i \cdot \vec{\varphi}}\,, \qquad X_1X_2X_3=1\,,\nn\\
&&\vec{a}_1=(\ft2{\sqrt6},\sqrt2)\,, \qquad 
\vec{a}_2=(\ft2{\sqrt6},-\sqrt2)\,, \qquad
\vec{a}_3=(-\ft4{\sqrt6},0)\,,
\eea
and the field strengths are defined as $F^i_\2=dA^i_\1$.
The equations of motion are
\bea
R_{\sst{MN}}&=&\ft12\del_{\sst M}\vec{\varphi}\cdot\del_{\sst N}\vec{\varphi}
- \ft43g^2g_{\sst{MN}}\sum_iX_i^{-1}\nn\\
&&\quad + \ft12\sum_iX_i^{-2}(F_{\sst{MP}}^{\,i}F_{\sst N}^{\,i\;\sst P}
-\ft16(F_\2^i)^2g_{\sst{MN}})\,,\nn\\
\nabla_{\sst M}(X_i^{-2}F_i^{\sst{MN}})&=&
\ft14\epsilon^{\sst{NPQRS}}
F^j_{\sst{PQ}}F^k_{\sst{RS}}\,, \qquad i \neq j \neq k \neq i\,,\nn\\
\square\vec{\varphi}&=&\ft14\sum_i\vec{a}_iX_i^{-2}(F_\2^{\,i})^2
-2g^2\sum_i\vec{a}_iX_i^{-1}.
\eea
The supersymmetry transformations for the fermions are given by
\bea
\delta\Psi_{\sst M}&=&{\big[}\nabla_{\sst M}-\ft{\rm i}2g\sum_iA_{\sst M}^i
+\ft16g\,\Gamma_{\sst M}\sum_iX_i
-\ft{\rm i}{48}(\Gamma_{\sst M}\Gamma^{\sst{AB}}
- 3\Gamma^{\sst{AB}}\Gamma_{\sst M})
\sum_iX_i^{-1}F_{\sst{AB}}^i{\big]}\epsilon\,,\nn\\
\delta\vec{\lambda}&=&{\big[}-\ft{\rm i}4\Gamma^{\sst M}\del_{\sst M}
\vec{\varphi}+\ft1{16}\Gamma^{\sst{AB}}\sum_i\vec{a}_iX_i^{-1}
F^i_{\sst{AB}} - \ft{\rm i}4g\sum_i\vec{a}_iX_i{\big]}\epsilon\,.
\eea

\subsection{The solution}
We use the following pp-wave metric ansatz 
\be
ds_{\sst D}=e^{2A}(-4dx^+dx^- + H(dx^+)^2 + dz_a^2) + e^{2B}dr^2\,, 
\qquad a=1,2,\cdots, D-3,
\ee
in arbitrary dimensions. The functions $A$ and $B$ depend on $r$ only while
$H$ depends on $x^+,\,z_a$ and $r$ coordinates.
If we set $H=0$ the pp-waves reduce to AdS-domain wall solutions \cite{cglp}.
It is natural to choose the following vielbein basis
\bea
e^+=e^{A}dx^+\,, \quad e^-=e^{A}(-2dx^- + \ft12Hdx^+)\,, \quad
e^a=e^{A}dz^a\,, \quad e^r=e^{B}dr
\eea
such that we have $ds^2=2e^+ e^- + e^a e^a + e^r e^r$\,.
The vielbein components of the spin connections are
\bea
\omega_{-r}&=&A'e^{-B}e^+\,, \quad \omega_{+a}=\ft12e^{-A}\del_a H\, e^+\,,\nn\\
\omega_{+r}&=&A'e^{-B}e^- + \ft12H'e^{-B}e^+\,, \quad \omega_{ar}=A'e^{-B}e^a\,.
\eea
where the prime denotes the derivative with respect to $r$. 
Note that for the metric in this basis we have $\eta_{+-}=1$ and
$\eta_{++}=\eta_{--}=0$. The derivatives are always with respect to 
the curved metric. The vielbein components of the Ricci tensor in 
$D$-dimensions are given by
\bea
R_{++}&=&-\ft12e^{-2B}\,{\big[}H''+H'((D-1)A'-B')]-\ft12e^{-2A}
\sum_a \del_a\del_a H = -\ft12\,\square H\,,\nn\\
R_{+-}&=&-e^{-2B}\,[A''+A'((D-1)A'-B')]\,, \qquad 
R_{ab}=R_{+-}\,\delta_{ab}\,,\nn\\
R_{rr}&=&-(D-1)e^{-2B}\,[A''+A'(A'-B')]\,.
\eea
It is straightforward to verify that in five dimensions the following
\bea
e^{2A}&=&(gr)^2\,[H_1H_2H_3]^{1/3}\,, \qquad H_i=1+\fft{\ell_i^2}{r^2}\,,\nn\\
e^{2B}&=&\fft1{(gr)^2\,[H_1H_2H_3]^{2/3}}\,, \qquad
X_i=H_i^{-1}[H_1H_2H_3]^{1/3}\,,\nn\\
A^i_\1&=&g^{-1}S_i(1-H_i^{-1})\,dx^+ \label{sol5}
\eea
satisfy the equations of motion with $H(x^+,r,z_a)$ obeying the equation
\be
H''+(4A'-B')H'+e^{-2(A-B)}\sum_a\del_a\del_a H
+ \fft{4g^2}{r^2}e^{-6A}\sum_i S_i^2 \ell_i^4H_i^{-2}=0\,.\label{eqH4}
\ee
Here the $S_i$ are functions of $x^+$.

\subsection{Standard supersymmetry}
The Killing spinor equations following from the fermionic transformations 
are given by
\bea
&&{\big[}\del_+ + \ft12A'e^{A-B}(\Gamma_+ + \ft12H\Gamma_-)
(\Gamma_r+1)-\ft14e^{A-B}H'\, \Gamma_r\, \Gamma_-\nn\\
&&\qquad - \ft14(\del_1 H\, \Gamma_1 + \del_2 H\, \Gamma_2)\Gamma_-
-\ft{\rm i}2{\Big(}\sum_i S_i(1-H_i^{-1}){\Big)}(\Gamma_r+1)\nn\\
&&\qquad + \fft{\rm i}{6r^2}{\Big(}\sum_i S_i\ell_i^2H_i^{-1}{\Big)}
\Gamma_r\, \Gamma_+\, \Gamma_-{\big]}\epsilon=0\,,\nn
\eea
\bea
&&{\big[}\del_- - A'e^{A-B}\,\Gamma_-(\Gamma_r+1){\big]}
\epsilon=0\,,\nn\\
&&{\big[}\del_a + \ft12A'e^{A-B}\,\Gamma_a(\Gamma_r+1)
-\fft{\rm i}{6r^2}{\Big(}\sum_i S_i\ell_i^2H_i^{-1}{\Big)}
\Gamma_a\, \Gamma_r\, \Gamma_-{\big]}\epsilon=0\,,\nn\\
&&{\big[}\del_r + \fft1{6r}{\Big(}\sum_iH_i^{-1}{\Big)}\Gamma_r
+\fft{\rm i}{3r}ge^{-3A}{\Big(}\sum_i S_i\ell_i^2H_i^{-1}{\Big)}
\Gamma_-{\big]}\epsilon=0\,,\nn\\
&&{\big[}{\rm i}e^{3A}{\Big(}\sum_ia_{1i}H_i^{-1}{\Big)}
(\Gamma_r+1)+g{\Big(}\sum_ia_{1i} S_i\ell_i^2H_i^{-1}{\Big)}
\Gamma_r\, \Gamma_-{\big]}\epsilon=0\,,\nn\\
&&{\big[}{\rm i}e^{3A}{\Big(}\sum_ia_{2i}H_i^{-1}{\Big)}
(\Gamma_r+1)+g{\Big(}\sum_ia_{2i} S_i\ell_i^2H_i^{-1}{\Big)}
\Gamma_r\, \Gamma_-{\big]}\epsilon=0\,,
\eea
where we have $\Gamma_+^2=\Gamma_-^2=0$ and $\{\Gamma_+\,,\Gamma_-\}=2$.
To arrive at these equations we have made use of the solution
(\ref{sol5}). The above Killing spinor equations have the solution
\be
\epsilon=r^{1/2}\,[H_1H_2H_3]^{\fft1{12}}\epsilon_0
\ee
where $\epsilon_0$ is a constant spinor satisfying $(\Gamma_r+1)\epsilon_0=0$ 
and $\Gamma_-\epsilon_0=0$. The solution therefore preserves 
$\ft14$ of the supersymmetry. The Killing spinor for the $\ft14$ 
supersymmetry exist for arbitrary solutions to eq.(\ref{eqH4}).

\subsection{Supernumerary supersymmetry}
To investigate the supernumerary supersymmetry we use the less
restrictive projection condition
\be
(\Gamma_r+1)\epsilon={\im} f\, \Gamma_-\epsilon
\ee
where the function $f=f(x^+,r,z_a)$ is to be determined. Making use of this 
projection in the Killing spinor equations they become
\bea
&&[\del_+ + \ft{\im}2(A'e^{A-B}f-\fft1{3r^2}{\cal M})
\Gamma_+\,\Gamma_- + \fft1{4r^2}(2f{\cal M}-r^2e^{A-B}H')\Gamma_-\nn\\
&&\qquad \qquad -\ft14(\del_1 H\, \Gamma_1
+ \del_2 H\, \Gamma_2)\Gamma_-]\epsilon=0\,,\nn\\
&&[\del_a + \ft{\im}2(A'e^{A-B}f-\fft1{3r^2}{\cal M})\Gamma_a\,\Gamma_-]
\epsilon=0\,, \qquad \del_-\epsilon=0\,,\nn\\
&&{\Big[}\del_r+\fft{\im}{6r}{\Big(}f\sum_i H_i^{-1} 
+ 2ge^{-3A}{\cal M}{\Big)}\Gamma_- 
- \fft1{6r}\sum_i H_i^{-1}{\Big]}\epsilon=0\,,\nn\\
&&{\Big[}e^{3A}f\sum_i a_{b i}H_i^{-1}-g\sum_i 
a_{b i}S_i\ell_i^2H_i^{-1}{\Big]}\Gamma_-\epsilon=0\,, \qquad b=1,2
\label{pkilling5}
\eea
where ${\cal M}\equiv \sum_iS_i\ell_i^2 H_i^{-1}$.
We analyse these equations by calculating the integrability 
conditions $[\del_{\sst M}\,, \del_{\sst N}]\epsilon=0$ among them.
The integrability $[\del_a\,,\del_r]\epsilon=0$ yields a solution for $f$ 
with the requirement $\del_a f=0$. We have
\be
f=\fft1{3r^2A'}\,e^{-(A-B)}({\cal M} + 3r^2U(x^+))\,,
\ee
where the function $U$ is in general complex. From the integrability
$[\del_+\,,\del_a]\epsilon=0$ we obtain an equation for $U$ after imposing
some restrictions on the pp-wave function $H(x^+,r,z_a)$. 
The result is
\bea
&&\del_a H'=0\,, \qquad \del_a\del_b H=0 \qquad 
\mbox{for}\qquad a\neq b\,,\nn\\
&&{\im}\fft{dU}{dx^+}+U^2+\ft12\del_a\del_a H=0\,, \qquad a=1,2\,.\label{U5}
\eea
The equation for $U$ then requires $\del_1\del_1 H = \del_2\del_2 H$.
Investigating the pair of equations given in the last line of 
eqs.(\ref{pkilling5}) we find that they are satisfied provided that 
the functions $S_i$ and $U$ satisfy two equations among them.
We present the solutions in terms of $S_3$ and $U$. They are given by
\be
S_1=\ell_1^{-2}{\big(}S_3\ell_3^2-(\ell_1^2-\ell_3^2)U{\big)}\,,\qquad
S_2=\ell_2^{-2}{\big(}S_3\ell_3^2-(\ell_2^2-\ell_3^2)U{\big)}\,.\label{5S12}
\ee
In order to analyse the final integrability $[\del_+\,,\del_r]\epsilon=0$
we need to make use of the solution for $H$.
Taking into account the conditions on $H$ given above the solution is
given by
\bea
g^4H(x^+,r,z_a)&=&\ft12c\, g^4(z_1^2 + z_2^2) 
+ \ft12|\epsilon_{ijk}|\,K_{ijk}(x^+,r)\,,\nn\\
K_{ijk}(x^+,r)&=&-\fft{S_i^2\ell_i^4}{(\ell_i^2-\ell_j^2)
(\ell_i^2-\ell_k^2)(r^2+\ell_i^2)}\nn\\
&& + \fft1{2(\ell_i^2-\ell_j^2)^2
(\ell_i^2-\ell_k^2)^2}{\Big[}(bg^4 + c\, \ell_i^2)
(\ell_i^2-\ell_j^2)(\ell_i^2-\ell_k^2)\\
&& + 2S_i^2\ell_i^4(2\ell_i^2-\ell_j^2-\ell_k^2)
- 2S_j^2\ell_j^4(\ell_i^2-\ell_k^2)
-2S_k^2\ell_k^4(\ell_i^2-\ell_j^2){\Big]}\ln(r^2+\ell_i^2)\,,\nn
\eea
where $b=b(x^+)$ and $c=c(x^+)$.
Then $[\del_+\,,\del_r]\epsilon=0$ yields an equation for $S_3$ given by
\be
{\im}\fft{dS_3}{dx^+}-(2\ell_3)^{-2}[bg^4 + c\,\ell_3^2 + 2U(U(\ell_1^2+\ell_2^2)
-2\ell_3^2(2S_3+U))]=0\,.\label{S5}
\ee
We proceed next by making use of the information that $S_i,\,b$ and $c$ are 
real functions. Eqs.(\ref{5S12}) implies that $U$ must also be real. 
This has the consequence in eq.(\ref{U5}) that $U$ and $c$ 
must be constants with $c$ being given by $c=-2U^2$\,.
Eqs.(\ref{S5}) and (\ref{5S12}) in turn implies that $S_i$ 
and $b$ must also be constants. Eliminating $U$ from eqs.(\ref{5S12}) 
and setting $S_i=\mu_i$ we obtain
\be
\epsilon_{ijk}\,\mu_i\ell_i^2(\ell_j^2-\ell_k^2)=0\,. \label{chargecond5}
\ee
Without loss of generality we solve for $\mu_1$ in terms of the other two charges.
The function $H$ which gives $\ft12$ supersymmetric pp-wave is given by
\bea
\mu_1&=&\fft{\mu_2\ell_2^2(\ell_1^2-\ell_3^2)-\mu_3\ell_3^2(\ell_1^2-\ell_2^2)}
{\ell_1^2(\ell_2^2-\ell_3^2)}\,,\nn
\eea
\bea
b&=&-\fft{2(\mu_2\ell_2^2-\mu_3\ell_3^2)
{\big(}\mu_2\ell_2^4-\mu_3\ell_3^4-3(\mu_2-\mu_3)\ell_2^2\ell_3^2
+\ell_1^2(\mu_2\ell_2^2-\mu_3\ell_3^2){\big)}}
{g^4(\ell_2^2-\ell_3^2)^2}\,,\nn\\
c&=&-\fft{2(\mu_2\ell_2^2-\mu_3\ell_3^2)^2}{(\ell_2^2-\ell_3^2)^2}\,,\nn\\
H&=&\ft12c\,(z_1^2+z_2^2)-f^2\,,\nn\\
f&=&-\fft{(\mu_2\ell_2^2-\mu_3\ell_3^2)r^2+(\mu_2-\mu_3)\ell_2^2\ell_3^2}
{g^2(\ell_2^2-\ell_3^2)r^3[H_1H_2H_3]^{1/2}}\,.
\eea
We next calculate the Killing spinor.
The projected Killing spinor equations become
\bea
&&[\del_+ - \ft1{2\sqrt2}(-c)^{1/2}({\rm i}\,\Gamma_+ - f)\Gamma_- 
- \ft14c\,(z_1\Gamma_1+z_2\Gamma_2)\Gamma_-]\epsilon=0\,,\nn\\
&&[\del_a - \ft{\rm i}{2\sqrt2}(-c)^{1/2}\,\Gamma_a\,\Gamma_-]
\epsilon=0\,,\qquad \del_-\epsilon=0\,,\nn\\
&&{\big[}\del_r - \ft{\rm i}2f'\,\Gamma_- 
- \fft1{6r}\sum_iH_i^{-1}{\big]}\epsilon=0\,.
\eea
The Killing spinor is easily obtained, given by
\bea
\epsilon&=&r^{1/2}[H_1H_2H_3]^{\ft1{12}}
{\big(}1+\ft{\rm i}{2\sqrt2}(-c)^{1/2}(z_1\Gamma_1+z_2\Gamma_2)
\Gamma_-{\big)}(1+\ft{\rm i}2f\,\Gamma_-)\eta\,,\nn\\
\fft{d\eta}{dx^+}&=&\ft{\rm i}{2\sqrt2}(-c)^{1/2}\,
\Gamma_+\,\Gamma_-\,\eta\,.
\eea
Solving for $\eta$ we have
\bea
\epsilon&=&r^{1/2}[H_1H_2H_3]^{\ft1{12}}
{\big(}1+\ft{\rm i}{2\sqrt2}(-c)^{1/2}(z_1\Gamma_1+z_2\Gamma_2)
\Gamma_-{\big)}(1+\ft{\rm i}2f\,\Gamma_-)\nn\\
&&\qquad \times {\big[}1-\ft12{\big(}1-e^{\ft{\im}{\sqrt2}
(-c)^{1/2}x^+}{\big)}\Gamma_+\,\Gamma_-{\big]}\epsilon_0\,,
\eea
where $\epsilon_0$ is a constant spinor satisfying $(\Gamma_r+1)\epsilon_0=0$\,.
The solution thus preserve $\ft12$ of the supersymmetry.
Note that if we set $\mu_i\ell_i^2=\mu$ (which is consistent with 
eq.(\ref{chargecond5})) we obtain $b=c=0$.

		To conclude, demanding supernumerary supersymmetry puts very strong 
restrictions on the pp-waves with the functions $S,\,b$ and $c$ (and $U$)
which initially all being functions of $x^+$ reduce now to constants.
This is not the case for minimal gauged supergravity where supernumerary 
supersymmetry does allow the various functions to have $x^+$ dependence.

\section{PP-waves in four dimensions}
In this section we consider a subsector of the $SO(8)$ gauged supergravity
where the bosonic fields comprises the metric, four commuting $U(1)$ 
gauge potentials and three dilatons. The Lagrangian describing this 
set of fields is \cite{duffliu}
\be
e^{-1}{\cal L}_4=R-\ft12(\del\vec{\varphi})^2
-\ft14\sum_iX_i^{-2}(F_\2^i)^2-V\,,
\ee
where $\vec{\varphi}=(\varphi_1,\varphi_2,\varphi_3)$, and
\bea
&&X_i=e^{-\ft12\vec{a}_i \cdot \vec{\varphi}}\,, \qquad X_1X_2X_3X_4=1\,,\nn\\
&&\vec{a}_1=(1,1,1)\,, \quad 
\vec{a}_2=(1,-1,-1)\,, \quad
\vec{a}_3=(-1,1,-1)\,, \quad \vec{a}_3=(-1,-1,1)\,.
\eea
The field strengths are defined as $F^i_\2=dA^i_\1$ and the
potential is given by
\be
V=-4g^2\sum_{i<\,j}X_iX_j=-8g^2\sum_{i=1}^3\cosh\varphi_i\,.
\ee
The ${\cal N}=8$ supersymmetry transformations in this bosonic 
background were also presented in \cite{duffliu}. 
They are given by
\bea
\delta\Psi_{\sst M}^i&=&\nabla_{\sst M}\epsilon^{(i)} + \sum_j{\big[}
-{\rm i}g\,\Omega_{ij}A_{\sst M}^j 
+ \ft{\rm i}8\Omega_{ij}X_j^{-1}F_{\sst{AB}}^j
\Gamma^{\sst{AB}}\Gamma_{\sst M} + \ft14gX_j\Gamma_{\sst M}{\big]}
\epsilon^{(i)},\nn\\
\delta\lambda^{ij}&=&{\big[}\ft{\rm i}{\sqrt2}\Gamma^{\sst M}
\del_{\sst M}\phi^{ij} - \ft1{2\sqrt2}\sum_k\Omega_{jk}X_k^{-1}
F^k_{\sst{AB}}\Gamma^{\sst{AB}} + {\rm i}\sqrt2g\sum_{k,m}f_{ijk}
\Omega_{km}X_m{\big]}\epsilon^{(i)},
\eea
where we have rewritten them by introducing complex fermions
$\Psi^i_{\sst M}=\Psi^i_{1\sst M}+{\rm i}\Psi^i_{2\sst M}$, etc 
and made the substitutions $g\rightarrow \sqrt2g$ and 
$A^i_\1\rightarrow -\ft1{2\sqrt2}A^i_\1$\,.
Note that $i \neq j$ in the spin $1/2$ transformations. 
The three dilatons are given by the following identifications
\be
\varphi_1=\phi^{12}=\phi^{34}\,,\qquad
\varphi_2=\phi^{13}=\phi^{24}\,,\qquad
\varphi_3=\phi^{14}=\phi^{23}\,,
\ee
and note also that $\phi^{ij}=\phi^{ji}$.  
The function $f_{ijk}$ is defined as
\be
f_{ijk} = \left\{ \begin{array}{r@{\quad\mbox{for}\quad}l}
          \left|\epsilon_{ijk}\right| & i,j,k \ne 1\,,\\
          \delta_{jk} & i=1\,,\\
          \delta_{ik} & j=1\,,
          \end{array} \right.
\ee
and the matrix $\Omega$ is given by
\be
\Omega=\fft12 \left( \begin{array}{rrrr}
       1 & 1 & 1 & 1\\
       1 & 1 & -1 & -1\\
       1 & -1 & 1 & -1\\
       1 & -1 & -1 & 1
       \end{array} \right)
\ee

\subsection{The solution}
The four charge pp-wave is given by
\be
ds_4^2=e^{2A}(-4dx^+dx^- + H(dx^+)^2 + dz^2) + e^{2B}dr^2,
\ee
where
\bea
e^{2A}&=&(gr)^4\,[H_1H_2H_3H_4]^{1/2}\,, \qquad H_i=1+\fft{\ell_i^2}{r^2}\,,\nn\\
e^{2B}&=&\fft1{(gr)^2\,[H_1H_2H_3H_4]^{1/2}}\,, \qquad
X_i=H_i^{-1}[H_1H_2H_3H_4]^{1/4}\,,\nn\\
A^i_\1&=&g^{-1}S_i(1-H_i^{-1})\,dx^+
\eea
and $S_i=S_i(x^+)$. The function $H(x^+,r,z)$ satisfies the equation
\be
H''+(3A'-B')H'+e^{-2(A-B)}\del_z\del_zH
+ \fft{4g^2}{r^2}e^{-4A}\sum_i S_i^2\ell_i^4H_i^{-2}=0\,.
\ee
The solution to this equation is similar to the solution in $D=5$.
The four charged pp-wave can be specialised to one, two and three 
active charges respectively.

\subsection{Supersymmetry}
The ${\cal N}=8$ supersymmetry have four different sectors.
We begin by analysing the Killing spinor equations for the sector 
$\epsilon^\1$. The supersymmetry transformations are given by
\bea
\delta\Psi^1_{\sst M}&=&\nabla_{\sst M}\epsilon^\1
+\sum_{i=1}^4{\big[}-\ft{\rm i}2g
A^i_{\sst M}+\ft{\rm i}{16}X_i^{-1}F^{\,i}_{\sst{AB}}
\Gamma^{\sst{AB}}\Gamma_{\sst M}+\ft14gX_i\Gamma_{\sst M}{\big]}
\epsilon^\1,\nn\\
\delta\lambda^{12}&=&{\big[}\ft{\rm i}{\sqrt2}\Gamma^{\sst M}
\del_{\sst M}\varphi_1-\ft1{4\sqrt2}\Gamma^{\sst{AB}}
(X_1^{-1}F^1_{\sst{AB}}+X_2^{-1}F^2_{\sst{AB}}-X_3^{-1}F^3_{\sst{AB}}
-X_4^{-1}F^4_{\sst{AB}})\nn\\
&&\quad +\ft{\rm i}{\sqrt2}g(X_1+X_2-X_3-X_4){\big]}\epsilon^\1,\nn\\
\delta\lambda^{13}&=&{\big[}\ft{\rm i}{\sqrt2}\Gamma^{\sst M}
\del_{\sst M}\varphi_2-\ft1{4\sqrt2}\Gamma^{\sst{AB}}
(X_1^{-1}F^1_{\sst{AB}}-X_2^{-1}F^2_{\sst{AB}}+X_3^{-1}F^3_{\sst{AB}}
-X_4^{-1}F^4_{\sst{AB}})\nn\\
&&\quad +\ft{\rm i}{\sqrt2}g(X_1-X_2+X_3-X_4){\big]}\epsilon^\1,\nn\\
\delta\lambda^{14}&=&{\big[}\ft{\rm i}{\sqrt2}\Gamma^{\sst M}
\del_{\sst M}\varphi_3-\ft1{4\sqrt2}\Gamma^{\sst{AB}}
(X_1^{-1}F^1_{\sst{AB}}-X_2^{-1}F^2_{\sst{AB}}-X_3^{-1}F^3_{\sst{AB}}
+X_4^{-1}F^4_{\sst{AB}})\nn\\
&&\quad +\ft{\rm i}{\sqrt2}g(X_1-X_2-X_3+X_4){\big]}\epsilon^\1.
\eea
The Killing spinor equations are readily written down and take the form
\bea
&&{\big[}\del_+ + \ft12A'e^{A-B}(\Gamma_+ + \ft12H\Gamma_-)
(\Gamma_r+1)-\ft14H'e^{A-B}\,\Gamma_r\, \Gamma_-
-\ft14\del_z H\, \Gamma_z\, \Gamma_-\nn\\
&&\qquad - \ft{\rm i}2{\Big(}\sum_{i=1}^4S_i(1-H_i^{-1}){\Big)}
(\Gamma_r+1)
+ \fft{\rm i}{4r^2}{\Big(}\sum_i{\cal M}_i{\Big)}
\Gamma_r\, \Gamma_+\, \Gamma_-{\big]}\epsilon^\1=0\,,\nn\\
&&[\del_- - A'e^{A-B}\,\Gamma_-(\Gamma_r+1)]\epsilon^\1=0\,,\nn\\
&&{\big[}\del_z + \ft12A'e^{A-B}\,\Gamma_z(\Gamma_r+1)
-\fft{\im}{4r^2}{\Big(}\sum_i{\cal M}_i{\Big)}
\Gamma_z\, \Gamma_r\, \Gamma_-{\big]}\epsilon^\1=0\,,\nn\\
&&{\big[}\del_r + \fft1{4r}{\Big(}\sum_iH_i^{-1}{\Big)}\Gamma_r
+\fft{\rm i}{4r}ge^{-2A}{\Big(}\sum_i{\cal M}_i{\Big)}\Gamma_- 
{\big]}\epsilon^\1=0\,,
\eea
\bea
&&[{\im}g(X_1+X_2-X_3-X_4)(\Gamma_r+1)
+\fft1{r^2}e^{-A}({\cal M}_1+{\cal M}_2-{\cal M}_3
-{\cal M}_4)\Gamma_r\, \Gamma_-]\epsilon^\1=0\,,\nn\\
&&[{\im}g(X_1-X_2+X_3-X_4)(\Gamma_r+1)
+\fft1{r^2}e^{-A}({\cal M}_1-{\cal M}_2+{\cal M}_3
-{\cal M}_4)\Gamma_r\, \Gamma_-]\epsilon^\1=0\,,\nn\\
&&[{\im}g(X_1-X_2-X_3+X_4)(\Gamma_r+1)
+\fft1{r^2}e^{-A}({\cal M}_1-{\cal M}_2-{\cal M}_3
+{\cal M}_4)\Gamma_r\, \Gamma_-]\epsilon^\1=0\,,\nn
\eea
where we have defined ${\cal M}_i\equiv S_i\ell_i^2H_i^{-1}$.
These equations have the solution
\be
\epsilon^\1=r[H_1H_2H_3H_4]^{\ft18}\,\epsilon^\1_0
\ee
where $\epsilon^\1_0$ is a constant spinor satisfying
$(\Gamma_r+1)\epsilon^\1_0=0=\Gamma_-\epsilon^\1_0$.
Thus $\ft14$ of the supersymmetry of the $\epsilon^\1$ sector is 
preserved (standard supersymmetry). It is easy to see that the same
amount of supersymmetry is preserved simultaneously in the other 
sectors. The pp-wave therefore preserves overall $\ft14$ of 
the ${\cal N}=8$ supersymmetry. 

	Now let us examine whether the solution admits supernumerary 
supersymmetry. We again make use of the ansatz
\be
(\Gamma_r+1)\epsilon^\1={\rm i}f_1\,\Gamma_-\epsilon^\1.
\ee
A similar analysis of the integrability conditions among 
the projected Killing spinor equations as in five dimensions 
shows that the functions $S_i(x^+),\,U(x^+),\,b(x^+)$ and $c(x^+)$ 
must again be constants. 
In $D=4$ there are now two conditions that must be satisfied among 
the charges for there to be supernumerary supersymmetry. 
Hence the pp-wave solution will depend on just two charge parameters. 
The constraints among the charges are given by
\bea
\ell_1^2\ell_4^2(\ell_2^2-\ell_3^2)(\mu_1-\mu_4)
&=&\ell_2^2\ell_3^2(\ell_1^2-\ell_4^2)(\mu_2-\mu_3)\,,\nn\\
(\ell_2^2-\ell_3^2)(\mu_1\ell_1^2-\mu_4\ell_4^2)
&=&(\ell_1^2-\ell_4^2)(\mu_2\ell_2^2-\mu_3\ell_3^2)\,,
\eea
where we have set $S_i=\mu_i$. Solving for $\mu_1$ and $\mu_2$ in
terms of the other two charges the function $H$ is given by
\bea
\mu_{\alpha}&=&\fft{\mu_3\ell_3^2(\ell_\alpha^2-\ell_4^2)-\mu_4\ell_4^2
(\ell_\alpha^2-\ell_3^2)}{\ell_\alpha^2(\ell_3^2-\ell_4^2)}\,, \qquad 
\alpha=1,2\,,\nn\\
b&=&-\fft{2(\mu_3\ell_3^2-\mu_4\ell_4^2)}{g^6(\ell_3^2-\ell_4^2)^2}
{\big[} \mu_3\ell_3^2(\ell_1^2+\ell_2^2+\ell_3^2-5\ell_4^2)
- \mu_4\ell_4^2(\ell_1^2+\ell_2^2-5\ell_3^2+\ell_4^2) {\big]}\,,\nn\\
c&=&-\fft{8(\mu_3\ell_3^2-\mu_4\ell_4^2)^2}{(\ell_3^2-\ell_4^2)^2}\,,\nn\\
H&=&\ft12c\, z^2-f_1^2\,,\nn\\
f_1&=&-\fft{(\mu_3\ell_3^2-\mu_4\ell_4^2)r^2+(\mu_3-\mu_4)\ell_3^2\ell_4^2}
{g^3(\ell_3^2-\ell_4^2)r^4[H_1H_2H_3H_4]^{1/2}}\,.
\eea
The projected Killing spinor equations are given by
\bea
&&[\del_+ - \ft1{2\sqrt2}(-c)^{1/2}({\rm i}\,\Gamma_+ - f_1)\Gamma_- 
- \ft14c\, z \Gamma_z\,\Gamma_-]\epsilon^\1=0\,,\nn\\
&&[\del_z - \ft{\im}{2\sqrt2}(-c)^{1/2}\,\Gamma_z\,\Gamma_-]
\epsilon^\1=0\,,\qquad \del_-\epsilon^\1=0\,,\nn\\
&&{\big[}\del_r - \ft{\rm i}2f'_1\,\Gamma_- 
- \fft1{4r}\sum_iH_i^{-1}{\big]}\epsilon^\1=0\,.
\eea
The solution for the Killing spinor is
\bea
\epsilon^\1&=&r[H_1H_2H_3H_4]^{\ft18}
{\big(}1+\ft{\rm i}{2\sqrt2}(-c)^{1/2}z\,\Gamma_z\,
\Gamma_-{\big)}(1+\ft{\rm i}2 f_1\,\Gamma_-)\nn\\
&&\qquad \times {\big[}1-\ft12{\big(}1-e^{\ft{\im}{\sqrt2}
(-c)^{1/2}x^+}{\big)}\Gamma_+\,\Gamma_-{\big]}\epsilon_0^\1,
\eea
where $\epsilon_0^\1$ is a constant spinor satisfying 
$(\Gamma_r+1)\epsilon_0^\1=0$\,. The pp-wave with $H$ given
above therefore preserves $\ft12$ of the supersymmetry of 
the $\epsilon^\1$ sector. Consider next the remaining sectors. 
For this we use the ansatz
$A_\1^i=g^{-1}\eta_i\, \mu_i(1-H_i^{-1})dx^+$.
To preserve $\ft12$ supersymmetry in the four respective 
sectors then requires the sign choices:
\be
\begin{array}{rrrrr}
1\,: & \eta_1=&\eta_2=&\eta_3=&\eta_4\\
2\,: & \eta_1=&\eta_2=&-\eta_3=&-\eta_4\\
3\,: & \eta_1=&-\eta_2=&\eta_3=&-\eta_4\\
4\,: & \eta_1=&-\eta_2=&-\eta_3=&\eta_4
\end{array}
\ee
Because of the difference in signs the four charge solution will preserve
$\ft12$ of the supersymmetry of just one sector and $\ft14$
of the supersymmetry of each of the remaining sectors.

		Although we have focused on solutions with four active charges one 
can easily also analyse the supersymmetry of solutions with one, two 
or three active charges. In the following table we present the overall
amount of the ${\cal N}=8$ supersymmetry preserved in the various cases.
\[\]
\begin{center}
\begin{tabular}{|c|c|c|}\hline
No. of active & Standard & Enhanced\\
charges& supersymmetry & supersymmetry\\ \hline\hline
1 & $\ft14$ & $\ft18+\ft18+\ft18+\ft18=\ft12$\\ \hline
2 & $\ft14$ & $\ft18+\ft18+\ft1{16}+\ft1{16}=\ft38$\\ \hline
3 & $\ft14$ & $\ft18+\ft1{16}+\ft1{16}+\ft1{16}=\ft5{16}$\\ \hline
4 & $\ft14$ & $\ft18+\ft1{16}+\ft1{16}+\ft1{16}=\ft5{16}$\\ \hline
\end{tabular}
\\[0.4cm]
\small{Table 1: Amount of ${\cal N}=8$ supersymmetry preserved by 1,\,2,\,3 
and 4 active charged pp-waves.}
\end{center}

\section{PP-waves in the Freedman-Schwarz model}
The Lagrangian describing the bosonic sector of the Freedman-Schwarz model
is \cite{freedman}
\bea
{\cal L}_4&=&R \ast\!\oneone - \ft12\ast\! d\phi \wedge d\phi
- \ft12e^{2\phi}\ast\! d\chi \wedge d\chi
+ 4(g_1^2+g_2^2)e^{\phi}\ast\!\oneone\nn\\
&&\quad -\ft12e^{-\phi}{\big(}\ast\! F^a_\2 \wedge F^a_\2 
+ \ast G^a_\2 \wedge G^a_\2 {\big)}
-\ft12\chi {\big(}F^a_\2 \wedge F^a_\2
+ G^a_\2 \wedge G^a_\2 {\big)}\,,
\eea
where
\bea
F^a_\2&=&dA^a_\1 - \ft1{\sqrt2}g_1\epsilon_{abc}A^b_\1 
\wedge A^c_\1\,, \qquad a=1,2,3,\nn\\
G^a_\2&=&dB^a_\1 - \ft1{\sqrt2}g_2\epsilon_{abc}B^b_\1 \wedge B^c_\1\,.
\eea
The supersymmetry transformations for the fermions are given by
\bea
\delta\Psi_{\sst M}&=&{\big[} \nabla_{\sst M}
- \ft{\rm i}{\sqrt2}g_1\alpha_1^aA^a_{\sst M} 
- \ft{\rm i}{\sqrt2}g_2\alpha_2^aB^a_{\sst M}
-\ft{\rm i}4e^{\phi}\,\Gamma_5\del_{\sst M}\chi\nn\\
&&\quad + \ft{\rm i}{8\sqrt2}e^{-\ft12\phi}(\alpha_1^aF^a_{\sst{AB}}
- {\rm i}\,\Gamma_5\alpha_2^aG^a_{\sst{AB}}) 
\Gamma^{\sst{AB}}\Gamma_{\sst M}
+ \ft12e^{\ft12\phi}(g_1 - {\rm i}g_2\Gamma_5)
\Gamma_{\sst M}{\big]}\epsilon\,,\nn\\
\delta\lambda&=&{\big[} \ft{\rm i}{\sqrt2}(\del_{\sst M}\phi
- {\rm i}e^{\phi}\,\Gamma_5\del_{\sst M}\chi)\Gamma^{\sst M}
+ \ft14e^{-\ft12\phi}(\alpha_1^aF^a_{\sst{AB}}
+ {\rm i}\,\Gamma_5\alpha_2^aG^a_{\sst{AB}})\Gamma^{\sst{AB}}\nn\\
&&\quad - {\rm i}\sqrt2\, e^{\ft12\phi}
(g_1+{\rm i}g_2\Gamma_5){\big]}\epsilon\,,
\eea
where $\Gamma_5={\rm i}\Gamma_0\Gamma_1\Gamma_2\Gamma_3$ such that
$\Gamma_5^2=1$. The $\alpha_1^a$ and $\alpha_2^a$ are two sets Pauli 
matrices. The gravitino, the dilatino and the (Majorana) spinor $\epsilon$ 
carry a suppressed indice which runs from one to four.
In the following we turn off two of the fields $F_{\sst{MN}}^a$ 
and $G_{\sst{MN}}^a$ each. For a vanishing axion $(\chi=0)$ the pp-wave 
in this theory is given by
\bea
ds^2&=&(gr)^2(-4dx^+dx^- + H(dx^+)^2 + dz^2) + dr^2\,,\nn\\
H&=&\ft12c\, z^2-\fft{b}{(gr)^2} - \fft{c\ln(gr)}{2g^2}
- \fft{g_2^2 S_1^2+g_1^2 S_2^2}{2g_1^2g_2^2(gr)^4}\,,\nn\\
\phi&=&-2\ln(gr)\,,\nn\\
A_\1&=&g_1^{-1}S_1(x^+){\big(}(gr)^{-2}-1{\big)}dx^+, \qquad
B_\1=g_2^{-1}S_2(x^+){\big(}(gr)^{-2}-1{\big)}dx^+,
\eea
where $g=(g_1^2+g_2^2)^{1/2}$. Now lets look at the supersymmetry of this solution.
The Killing spinor equations are given by
\bea
&&{\big[} \del_+ + \ft12g(\Gamma_+ + \ft12H\Gamma_-) 
(\Gamma_r+a) - \ft{\rm i}{\sqrt2}(S_1+S_2)
{\big(}(gr)^{-2}-1{\big)}\nn\\
&&\qquad - \ft14c\,z\Gamma_z\, \Gamma_- - \ft14grH'\,\Gamma_r\,\Gamma_-
- {\rm i}\Lambda\, \Gamma_-\,\Gamma_+\,\Gamma_r{\big]}\epsilon=0\,,\nn\\
&&{\big[}\del_- - g\Gamma_-(\Gamma_r+a){\big]}\epsilon=0\,,\nn\\
&&{\big[}\del_z + \ft12g\Gamma_z(\Gamma_r+a)
+ {\rm i}\Lambda\Gamma_z\,\Gamma_-\,\Gamma_r{\big]}\epsilon=0\,,\nn\\
&&{\big[}\del_r + \fft{{\rm i}\Lambda}{gr}\,\Gamma_-
+ \fft1{2gr}(g_1-{\rm i}g_2\Gamma_5)\Gamma_r{\big]}\epsilon=0\,,\nn\\
&&{\big[} (\Gamma_r+a)+ 2{\rm i}g^{-1}\Gamma_-\, \Lambda\, \Gamma_r 
{\big]}\epsilon=0\,,
\eea
where we have defined
\be
a=g^{-1}(g_1+{\rm i}g_2\Gamma_5) \qquad \mbox{and} \qquad
\Lambda=\fft{g_2S_1-{\rm i}g_1S_2\Gamma_5}
{2\sqrt2g_1g_2(g_1^2+g_2^2)^{1/2}r^2}\,.
\ee
It follows from these equations that to obtain the usual 
$\ft14$ supersymmetry for the pp-wave we need surprisingly 
to impose $g_2^2S_1=g_1^2S_2$\,. 
The Killing spinor can then be obtained and it is given by
\be
\epsilon=e^{-\ft{\rm i}{\sqrt2}\int (S_1+S_2)dx^+}r^{1/2}\epsilon_0\,,
\ee
where $\epsilon_0$ is a constant spinor satisfying the projections 
$(\Gamma_r+a)\epsilon_0=0=\Gamma_-\epsilon_0$\,. 
To investigate the supernumerary supersymmetry we use the 
projection condition
\be
(\Gamma_r+a)\epsilon={\rm i}f\,\Gamma_-\epsilon\,.\label{p5FS}
\ee
The projected Killing spinor equations are given by
\bea
&&[\del_+ +\ft{\im}{\sqrt2}(S_1+S_2)
+ {\im}\Gamma_+(\ft12g f-{\bar a}{\bar \Lambda})\Gamma_-
-\ft14c\, z\,\Gamma_z\, \Gamma_-\nn\\
&&\qquad -\fft1{2\sqrt2g_1g_2g^2r^2}(S_1g_2^2-S_2g_1^2)\Gamma_5
+(2\Lambda f-\ft14a\, grH')\Gamma_-]\epsilon=0\,,\nn\\
&&[\del_z + {\im}\Gamma_z(\ft12g f-{\bar a}{\bar \Lambda})\Gamma_-]
\epsilon=0\,,\qquad \del_-\epsilon=0\,,\nn\\
&&{\big[}\del_r + \fft{\im}{r}(g^{-1}\Lambda+\ft12{\bar a}f)\Gamma_-
- \fft1{2r}{\big]}\epsilon=0\,,\nn\\
&&[f-2g^{-1}{\bar a}{\bar \Lambda}]\Gamma_-\epsilon=0\,.\label{pFS}
\eea
Here $\bar a$ and $\bar \Lambda$ are just $a$ and $\Lambda$ 
but with $\Gamma_5$ replaced by $-\Gamma_5$.
We analyse these projected equations by calculating the integrability
conditions among them. The condition $[\del_z\,,\del_r]\epsilon=0$ 
requires $\del_z f=0$ and yields a solution for $f$ given by
\be
f=2g^{-1}{\bar a}{\bar \Lambda}+2g^{-1}U(x^+)\,.
\ee
The integrability $[\del_+\,,\del_z]\epsilon=0$ provides an equation for 
$U(x^+)$ which is given by
\be
{\im}\fft{dU}{dx^+}-2U^2-\ft14c=0\,.
\ee
From the last line of eqs.(\ref{pFS}) we have 
$f-2g^{-1}{\bar a}{\bar \Lambda}=0$. This equation forces
$U$ in the solution for $f$ to vanish. 
From the equation for $U$ we must in turn set $c=0$.
Considering next the integrability condition 
$[\del_+\,,\del_r]\epsilon=0$ we first note that
\be
2\Lambda f-\ft14a\,grH'=\fft{c\,r^2-4b}{8(gr)^2}(g_1-{\im}g_2\Gamma_5)\,.
\ee
It follows that the functions $S_1$ and $S_2$ must be constants. 
We need furthermore also to set $b=0$ (as well as imposing 
$g_2^2S_1=g_1^2S_2$). Letting $S_i=\mu_i$ the projected Killing 
spinor equations become
\bea
&&[\del_+ + \ft{\im}{\sqrt2}(\mu_1+\mu_2)]\epsilon=0\,, \qquad 
\del_-\epsilon=0\,, \qquad \del_z\epsilon=0\,,\nn\\
&&{\Big[}\del_r - \fft{\im}{2} \fft{g_1-{\im}g_2\Gamma_5}
{(g_1^2+g_2^2)^{1/2}}f'\, \Gamma_-
- \fft1{2r}{\Big]}\epsilon=0\,.
\eea
The Killing spinor solution is
\be
\epsilon=e^{-\ft{\rm i}{\sqrt2}(\mu_1+\mu_2)x^+}r^{1/2}
{\Big[} 1+\fft{\im}{2} \fft{g_1-{\im}g_2\Gamma_5}
{(g_1^2+g_2^2)^{1/2}}f\, \Gamma_- {\Big]} \epsilon_0\,,
\ee
where $\epsilon_0$ is a constant spinor.
Inserting the Killing spinor in the projection condition
(\ref{p5FS}) and using
\be
f=\fft{\mu_1}{\sqrt2 g_1^2 (g_1^2+g_2^2)^{1/2}\, r^2}
\ee
we obtain $(\Gamma_r+a)\epsilon_0=0$. Thus, the pp-wave preserves 
$\ft12$ of the supersymmetry with $H$ given by
\be
H = -f^2 = -\fft{\mu_1^2}{2g_1^4(g_1^2+g_2^2)r^4}\,.
\ee

\section{PP-waves in six dimensions}
In this section we investigate the supersymmetry of pp-waves in
Romans theory \cite{romans}. We use the conventions of \cite{llpbrane}.
We consider a subsector of the theory by truncating out the $2$-form potential 
and the $U(1)$ potential. The Lagrangian describing the remaining fields is
given by
\be
e^{-1}{\cal L}=R-\ft12(\del\varphi)^2-\ft14X^{-2}(F^a_\2)^2
+ 4g^2(X^2+\ft43X^{-2}-\ft19X^{-6})
\ee
where $X=e^{-\ft1{2\sqrt2}\varphi}$ and 
$F^a_\2=dA^a_\1-\ft1{\sqrt2}g\,\epsilon_{abc}A^b_\1\wedge A^c_\1$.\\
We have here set $g_1=g_2=-\sqrt2g$ in \cite{llpbrane}. 
The supersymmetry transformations are
\bea
\delta\Psi_{{\sst M}i}&=&[D_{\sst M} + \ft14g(X+\ft13X^{-3})
\Gamma_{\sst M}]\epsilon_i - \ft{\rm i}{16\sqrt2}
(\Gamma_{\sst M}\Gamma^{\sst{AB}}-2\Gamma^{\sst{AB}}\Gamma_{\sst M})
X^{-1}F_{\sst{AB}i}{}^j\epsilon_j\,,\nn\\
\delta\lambda_i&=&[-\ft1{2\sqrt2}\Gamma^{\sst M}\del_{\sst M}\varphi
- \ft12g(X-X^{-3})]\epsilon_i - \ft{\rm i}{8\sqrt2}\Gamma^{\sst{AB}}
X^{-1}F_{\sst{AB}i}{}^j\epsilon_j\,,
\eea
where $D_{\sst M}\epsilon_i=\nabla_{\sst M}\epsilon_i 
- \ft{\rm i}{\sqrt2}gA_{\sst{M}i}{}^j\epsilon_j$.
We obtain the pp-wave with two of the $SU(2)$ fields turned off.  
The solution is given by
\bea
e^{2A}&=&(gr)^{4/3}H_1^{1/2}\,, \qquad H_1=1+\fft{\ell_1^2}{r^2}\,,\nn\\
e^{2B}&=&\fft1{(gr)^2H_1^{3/2}}\,, \qquad e^{\sqrt2\varphi}=H_1\,,\nn\\
A^1_\1&=&g^{-1}S_1(x^+)(1-H_1^{-1})\,dx^+\,,
\eea
and the pp-wave function $H(x^+,r,z_a)$ satisfies the equation
\be
H''+(5A'-B')H'+e^{-2(A-B)}\sum_a \del_a\del_a H
+ \fft{4S_1^2\ell_1^4}{g^2r^6(gr)^{4/3}H_1^4}=0\,.
\ee
The Killing spinor equations are given by
\bea
&&[\del_+ + \ft12A'e^{A-B}(\Gamma_+ + \ft12H\Gamma_-)(\Gamma_r+1)
-\ft14e^{A-B}H'\, \Gamma_r\, \Gamma_- - \ft14\sum_a\del_a H\,
\Gamma_a\,\Gamma_-\nn\\
&&\qquad -\ft{\rm i}{\sqrt2}S_1(1-H_1^{-1})(\Gamma_r+1)
+\fft{\rm i}{4\sqrt2}\fft{S_1 \ell_1^2}{r^2H_1}
\Gamma_r\, \Gamma_+\, \Gamma_-]\epsilon=0\,,\nn\\
&&[\del_- - A'e^{A-B}\,\Gamma_-(\Gamma_r+1)]\epsilon=0\,,\nn\\
&&[\del_a + \ft12A'e^{A-B}\,\Gamma_a(\Gamma_r+1)
- \fft{\im}{4\sqrt2} \fft{S_1 \ell_1^2}{r^2H_1}
\Gamma_a\, \Gamma_r\, \Gamma_-]\epsilon=0\,,\nn\\
&&[\del_r + \fft{\ell_1^2+4r^2}{12r^3H_1}\Gamma_r
+ \fft{3\rm i}{4\sqrt2}\fft{S_1 \ell_1^2(gr)^{1/3}}{g^2r^4H_1^2}
\Gamma_-]\epsilon=0\,,\nn\\
&&[g(\Gamma_r+1)+\fft{\rm i}{\sqrt2}\fft{S_1}{(gr)^{2/3}H_1}
\Gamma_r\, \Gamma_-]\epsilon=0\,.
\eea
It is clear from these equations that the pp-waves preserve $\ft14$ of 
the supersymmetry but there is no supernumerary supersymmetry.

\section{PP-waves in seven dimensions}
In this section we consider gauged $D=7,{\cal N}=2$ supergravity where we retain 
only the metric, two $U(1)$ gauge potentials and two scalars. 
The other fields are consistently set to zero. 
This reduced set of fields are described by the Lagrangian
\be
e^{-1}{\cal L}=R-\ft12(\del\vec\varphi)^2
-\ft14\sum_{i=1}^2X_i^{-2}(F_\2^i)^2-V\,,
\ee
where 
\bea
&&X_i=e^{-\ft12\vec{a}_i\cdot\vec\varphi}\,, \quad 
\vec{a}_1={\big(}\sqrt2,\sqrt{\ft25}{\big)}\,, \quad
\vec{a}_2={\big(}-\sqrt2,\sqrt{\ft25}{\big)}\,,\nn\\
&&V=\ft12g^2(X_1^{-4}X_2^{-4}-8X_1X_2-4X_1^{-1}X_2^{-2}
-4X_1^{-2}X_2^{-1})\,.
\eea
The supersymmetry transformations are given by
\bea
\delta\psi_{\sst M}&=&[\nabla_{\sst M}+\ft14(X_1^{-1}F^1_{\sst{MN}}
\Gamma_{12}+X_2^{-1}F^2_{\sst{MN}}\Gamma_{34})\Gamma^{\sst N}
+\ft14gX_1^{-2}X_2^{-2}\,\Gamma_{\sst M}\nn\\
&&\quad +\ft14(X_1^{-1}\del_{\sst N}X_1+X_2^{-1}\del_{\sst N}X_2)
\Gamma_{\sst M}\Gamma^{\sst N}+\ft12g(A^1_{\sst M}\Gamma_{12}
+A^2_{\sst M}\Gamma_{34})]\epsilon\,,\\
\delta\lambda_1&=&[-\ft18(3X_1^{-1}\del_{\sst M}X_1+2X_2^{-1}
\del_{\sst M}X_2)\Gamma^{\sst M}-\ft1{16}X_1^{-1}F^1_{\sst{AB}}
\Gamma^{\sst{AB}}\,\Gamma_{12}+\ft14g(X_1-X_1^{-2}X_2^{-2})]\epsilon\,,\nn\\
\delta\lambda_2&=&[-\ft18(2X_1^{-1}\del_{\sst M}X_1+3X_2^{-1}
\del_{\sst M}X_2)\Gamma^{\sst M}-\ft1{16}X_2^{-1}F^2_{\sst{AB}}
\Gamma^{\sst{AB}}\,\Gamma_{34}+\ft14g(X_2-X_1^{-2}X_2^{-2})]\epsilon\,.\nn
\eea
For more details see \cite{liu}. The domain wall solution is given by
\bea
e^{2A}&=&(gr)[H_0^{1/2}H_1H_2]^{\ft15}\,, \qquad 
H_i=1+\fft{\ell_i^2}{r^2}\,,\nn\\
e^{2B}&=&\fft1{(gr)^2[H_0^{1/2}H_1H_2]^{\ft45}}\,, \quad
X_i=H_i^{-1}[H_0^{1/2}H_1H_2]^{\ft25}\,,
\eea
where $H_0=1+\ell_0^2/r^2$. The ansatz for the $1$-form potential is
\be
A^i_\1=g^{-1}S_i(1-H_i^{-1})\,dx^+
\ee
and the function $H(x^+,r,z_a)$ satisfy the equation
\be
H''+(6A'-B')H'+e^{-2(A-B)}\sum_a \del_a\del_a H
+ \fft{4g^2}{r^2}e^{-10A}\sum_i S_i^2\ell_i^4H_i^{-2}=0\,.
\ee
The Killing spinor equations are given by
\bea
&&{\big[}\del_+ + \fft1{4rH_0}e^{A-B}(\Gamma_+ + \ft12H\Gamma_-)
(\Gamma_r+1)-\ft14e^{A-B}H'\, \Gamma_r\, \Gamma_- 
- \ft14\sum_a \del_a H\, \Gamma_a\, \Gamma_-\nn\\
&&\qquad +\ft12{\big(}S_1(1-H_1^{-1})\Gamma_{12}+S_2(1-H_2^{-1})
\Gamma_{34}{\big)}(\Gamma_r+1){\big]}\epsilon=0\,,\nn\\
&&[\del_- - \fft1{2rH_0}e^{A-B}\,\Gamma_-(\Gamma_r+1)]\epsilon=0\,,\nn\\
&&[\del_a + \fft1{4rH_0}e^{A-B}\,\Gamma_a(\Gamma_r+1)]\epsilon=0\,,\nn\\
&&[\del_r - \ft1{2\sqrt{10}}\varphi'_2 + \fft1{4rH_0}\Gamma_r
-\fft1{2r}ge^{-5A}(S_1\ell_1^2H_1^{-1}\Gamma_{12}
+S_2\ell_2^2H_2^{-1}\Gamma_{34})\Gamma_-]\epsilon=0\,,\nn\\
&&[g(\ell_0^2-\ell_1^2)H_0^{-1} X_1(\Gamma_r+1)
+S_1\ell_1^2H_1^{-1}e^{-A}\,\Gamma_{12}\, \Gamma_r\, \Gamma_-]
\epsilon=0\,,\nn\\
&&[g(\ell_0^2-\ell_2^2)H_0^{-1} X_2(\Gamma_r+1)
+S_2\ell_2^2H_2^{-1}e^{-A}\,\Gamma_{34}\, \Gamma_r\, \Gamma_-]
\epsilon=0\,.
\eea
It is clear from these equations that the pp-waves have $\ft14$ supersymmetry
but no supernumerary supersymmetry.

\section*{Acknowledgments}
I am grateful to Hong L\"u and Chris Pope for helpful discussions.

\end{document}